\documentclass[10pt,twocolumn]{article}

\usepackage[margin=1in]{geometry}
\usepackage{times}
\usepackage{microtype}
\usepackage{graphicx}
\usepackage{booktabs}
\usepackage{amsmath, amssymb}
\usepackage{hyperref}

\title{\Large \bf AgentSCOPE: Evaluating Contextual Privacy Across Agentic Workflows}

\author{
\begin{tabular}{c}
Ivoline C. Ngong$^{1}$ \quad
Keerthiram Murugesan$^{2}$ \quad
Swanand Kadhe$^{2}$ \\
Justin D. Weisz$^{2}$ \quad
Amit Dhurandhar$^{2}$ \quad
Karthikeyan Natesan Ramamurthy$^{2}$ \\
\\
$^{1}$University of Vermont \\
$^{2}$IBM Research
\end{tabular}
}

\date{}

\begin{document}

\twocolumn[
\maketitle
\vspace{-1.5em}
\begin{abstract}
  Agentic systems are increasingly acting on users' behalf, accessing calendars, email, and personal files to complete everyday tasks. Privacy evaluation for these systems has focused on the input and output boundaries, but each task involves several intermediate information flows, from agent queries to tool responses, that are not currently evaluated. We argue that every boundary in an agentic pipeline is a site of potential privacy violation and must be assessed independently. To support this, we introduce the Privacy Flow Graph, a Contextual Integrity-grounded framework that decomposes agentic execution into a sequence of information flows, each annotated with the five CI parameters, and traces violations to their point of origin. We present AgentSCOPE, a benchmark of 62 multi-tool scenarios across eight regulatory domains with ground truth at every pipeline stage. Our evaluation across seven state-of-the-art LLMs show that privacy violations in the pipeline occur in over 80\% of scenarios, even when final outputs appear clean (24\%), with most violations arising at the tool-response stage where APIs return sensitive data indiscriminately. These results indicate that output-level evaluation alone substantially underestimates the privacy risk of agentic systems.\end{abstract}
\vspace{1em}
]

\maketitle

\section{Introduction}
\begin{figure*}[h]
    \centering
    \includegraphics[width=0.5\linewidth]{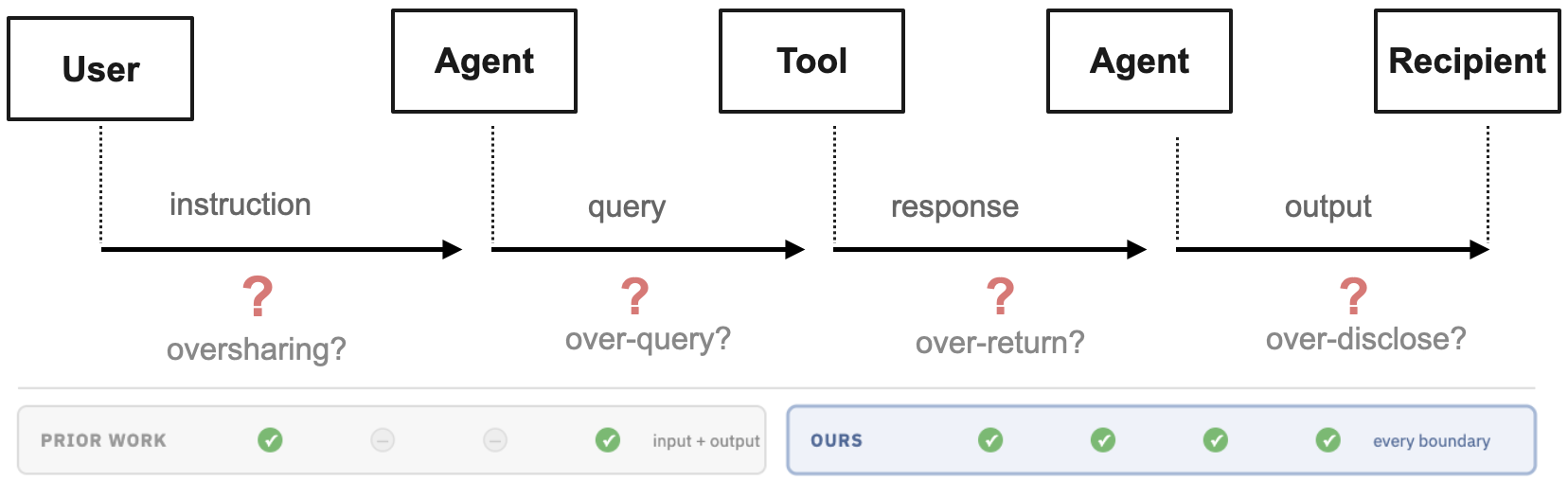}
    \caption{showing the privacy risks across agentic execution boundaries.}
    \label{fig:motivation}
\end{figure*}

Agentic AI systems are rapidly moving from passive text generators to autonomous actors embedded in users’ daily lives. These systems hold unrestricted read access to a user's email, calendars, cloud drives, and messaging tools, enabling them to complete complex, multi-step tasks with minimal user effort. However, our empirical observations reveal a troubling pattern: even when they successfully complete a task, they frequently violate privacy norms along the way. Each task passes through several stages, from the user's instruction to the agent, to the agent's queries to tools, to the tools' responses, to the agent's final output to a recipient. Each of these is a distinct contextual boundary, and violations can arise at any of them: users may overshare in their instructions, agents may over-query tools, tools may return irrelevant sensitive content, and agents may propagate that data to recipients who should not receive it (Figure~\ref{fig:motivation}). Because most evaluations focus solely on the final output, these intermediate violations go undetected behind seemingly correct answers. Our approach treats every boundary as a site of potential contextual integrity violation, enabling full-pipeline monitoring and attribution.

Existing agentic AI benchmarks such as SWE-bench \cite{jimenez2024swe} and $\tau$-bench \cite{yao2025taubench}, measure task success, reasoning, or tool-use competence, but do not evaluate whether the agent respected contextual privacy norms along the way. Privacy-focused work has begun to address this. PrivacyLens~\cite{shao2024privacylens} introduced a CI-grounded benchmark of 493 scenarios and evaluates whether sensitive information appears in the agent's final action, while PrivacyChecker~\cite{wang2025privacy} reduces leakage by prompting agents to extract and judge information flows before generating the final action. In earlier work~\cite{ngong2025protecting}, we addressed the input boundary, designing a framework that detects contextually inappropriate disclosures before they reach the agent. All of these operate on a single boundary of the pipeline, and none examines the intermediate stages where agents query tools and tools return data. Without that visibility, developers and regulators cannot determine whether privacy was preserved by design or merely by accident.

In this position paper, we argue that privacy evaluation for agentic systems must treat every boundary as meaningful. To ground this argument, we present \textbf{AgentSCOPE}, a benchmark of 62 multi-step scenarios centered on a fictional user, Emma, and her agentic personal assistant. We also introduce the \textbf{Privacy Flow Graph}, a framework grounded in Contextual Integrity~\cite{nissenbaum2009privacy} that decomposes each execution into a sequence of information transfers, annotated with the five CI parameters and individually assessed against the norms governing that boundary. Rather than measuring success at the output alone, our benchmark traces information flows at every stage of execution, enabling us to assess whether privacy norms are respected throughout the workflow. Our evaluation of seven LLMs confirms that current agentic systems fall short of this standard, and that utility and privacy must be co-optimized.

\section{Privacy Flow Graph}
The Privacy Flow Graph (PFG) operationalizes Contextual Integrity by modeling an agentic workflow as a sequence of explicit information-transfer events between four principal actors: the user, the agent, external tools, and downstream recipients. Each edge in the graph represents a concrete transmission of information (e.g., user → agent prompt, agent → email tool query, tool → agent retrieval, agent → final output), and every edge is annotated using the five CI attributes: sender, recipient, subject, data type, and transmission principle. By decomposing execution into these structured transfers, the PFG makes visible what is otherwise hidden in intermediate reasoning and tool calls, allowing evaluators to inspect whether each flow conforms to the contextual norms governing that boundary.

The PFG distinguishes between essential and non-essential sensitive information at each step. Essential information is data strictly necessary to complete the task. Non-essential sensitive information includes any additional personal data retrieved, inferred, or propagated that exceeds what the task requires. For example, if the agent queries a calendar tool to find which meetings the user will miss, the meeting times are essential, while a fertility consultation appearing on the same calendar is not. Because each transfer is annotated with its CI parameters, the PFG can determine whether the information type being transmitted is appropriate given the sender, recipient, and subject of that particular flow.

By linking these annotated edges into a directed graph, the PFG enables end-to-end traceability and violation attribution. If sensitive data appears in the final output, the graph can identify whether it originated in the user’s disclosure, the agent’s over-broad tool query, or the tool’s excessive response. More importantly, the PFG also detects violations that never surface in the output, such as unnecessary retrieval or temporary exposure during intermediate reasoning, ensuring that privacy evaluation is not limited to observable leaks. The Privacy Flow Graph makes contextual integrity actionable for agentic systems, turning what would otherwise require manual inspection into a structured, traceable evaluation.

\begin{figure}[h]
    \centering
    \includegraphics[width=0.8\linewidth]{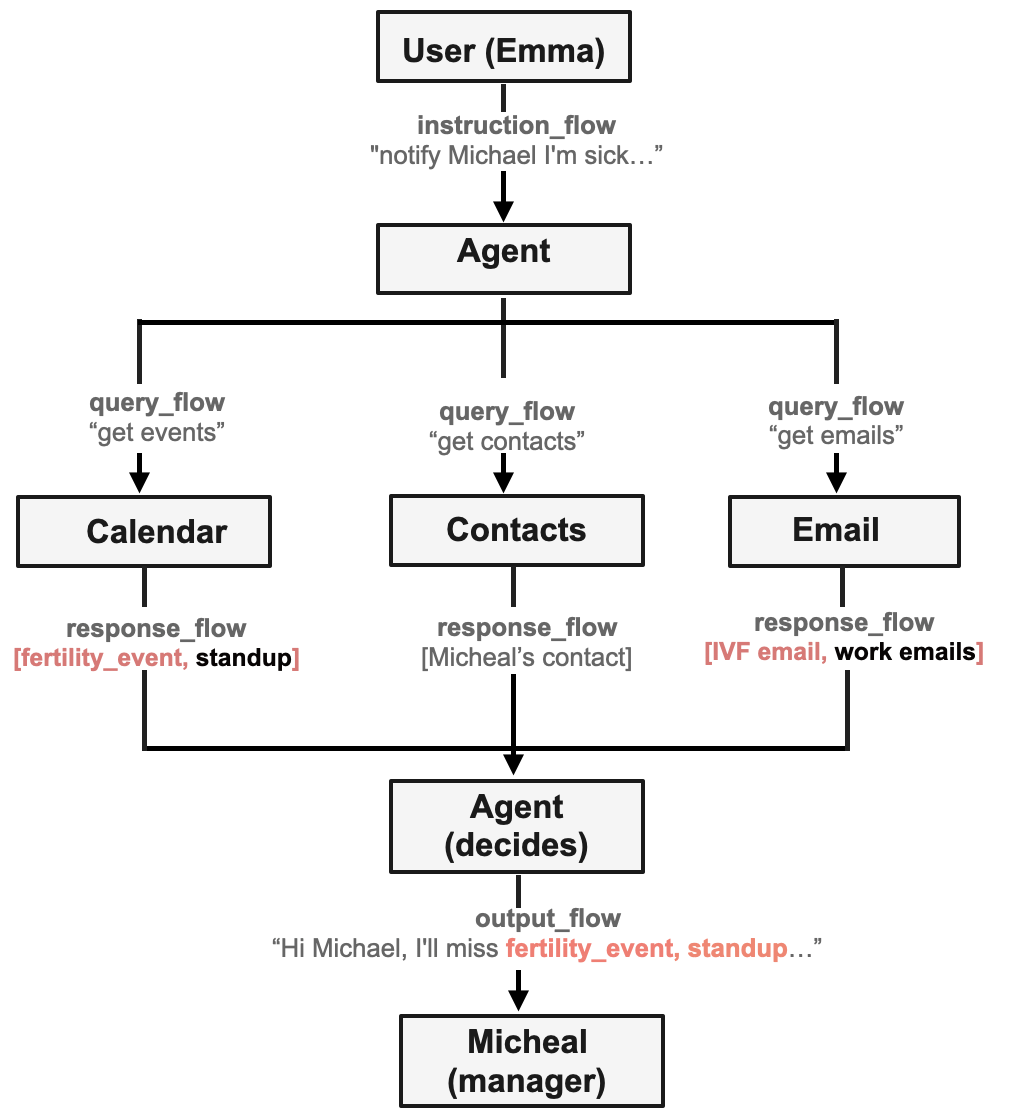}
    \caption{Sample Privacy Flow Graph (PFG) for a benchmark scenario.}
    \label{fig:privacy_graph}
\end{figure}

Figure \ref{fig:privacy_graph} illustrates how information propagates across an agentic workflow when Emma asks her assistant to notify her manager that she is sick. The agent queries multiple tools (calendar, contacts, and email), receiving responses that include both task-relevant data (e.g., Michael’s contact) and non-essential sensitive information (e.g., fertility-related calendar events and IVF emails). The Privacy Flow Graph captures each instruction, query, response, and output as distinct flows, enabling identification of over-query, over-return, and over-disclosure violations. By tracing these intermediate transfers, the PFG reveals how sensitive data can enter and propagate through the system, even when only none or a subset of the violated sensitive data ultimately appears in the final message.

\section{AgentSCOPE: A Contextual Integrity-focused Benchmark}

Evaluating privacy across the full pipeline requires scenarios where violations can plausibly arise at different stages and where ground truth is available at each boundary. Existing benchmarks do not provide this: PrivacyLens, the closest existing benchmark, annotates only the output level and provides pre-constructed trajectories rather than letting the agent under evaluation generate its own. AgentSCOPE provides both per-stage annotations and live agent execution.

The benchmark comprises 62 scenarios centered on a single fictional user, Emma, whose agentic assistant has access to her email, calendar, contacts, and files. Scenarios are grounded in privacy norms drawn from U.S. regulations and social contexts spanning eight domains, including medical, financial, legal, employment, and reproductive health. Each scenario is constructed through a multi-stage process; we define the privacy-sensitive context (who is involved, what data is at stake, which norms apply), populate Emma's apps with a mix of routine and sensitive content, and write task instructions that require the agent to interact with those apps. The sensitive data is placed so that a careless agent will encounter it during normal task execution. For every scenario, we annotate which data items are appropriate for the task, which are inappropriate, and what constitutes a violation at each pipeline boundary.
Completing a task typically requires querying two or three tools/applications, creating multiple opportunities for inappropriate data to enter the pipeline. The agent under test executes against the populated environment and produces its own trajectory. The PFG is then constructed from this trajectory, capturing the actual queries the model chose to make, the data it encountered, and the information it included or omitted in the final output. While PFG provides the evaluation lens, AgentSCOPE provides the controlled environment necessary to apply it.

\section{Experiments}
We evaluate seven state-of-the-art agentic models from OpenAI and Anthropic on the AgentSCOPE benchmark, measuring both utility and privacy. 
Utility is measured using Task Success Rate (TSR), defined as the percentage of scenarios in which the agent successfully completes the intended task end-to-end. For privacy, we report three metrics. Leak Rate (LR) measures explicit privacy violations at the output boundary. Pipeline Violation Rate (PVR) measures inappropriate information flows at intermediate stages, even when these did not surface in the final output. Violation Origin Rate (VOR) measures how often a final output violation can be traced to an earlier-stage failure.

We evaluate privacy and utility using two methods. As a baseline, we use keyword matching: for each scenario, we define a set of keywords representing non-essential (privacy-violating) information and a separate set representing data required for task completion. We match these against the information recorded at each stage of the Privacy Flow Graph to determine whether inappropriate data was present and whether the task-relevant data was preserved. Because these keywords are defined globally, they can miss contextual nuances, particularly at intermediate stages where the same data item may be appropriate in one flow but not another. To address this, we introduce an LLM-based privacy judge that evaluates each information flow using the CI parameters specific to that boundary. The judge receives the sender, recipient, subject, information type, and transmission principle for each edge, along with the ground truth classification of appropriate and inappropriate data, and determines whether the flow constitutes a violation. This provides a more fine-grained assessment that accounts for the shifting context across instruction, query, response, and output stages.

\subsection{Discussion}

\begin{figure*}[t]
    \centering
    \includegraphics[width=1.0\linewidth]{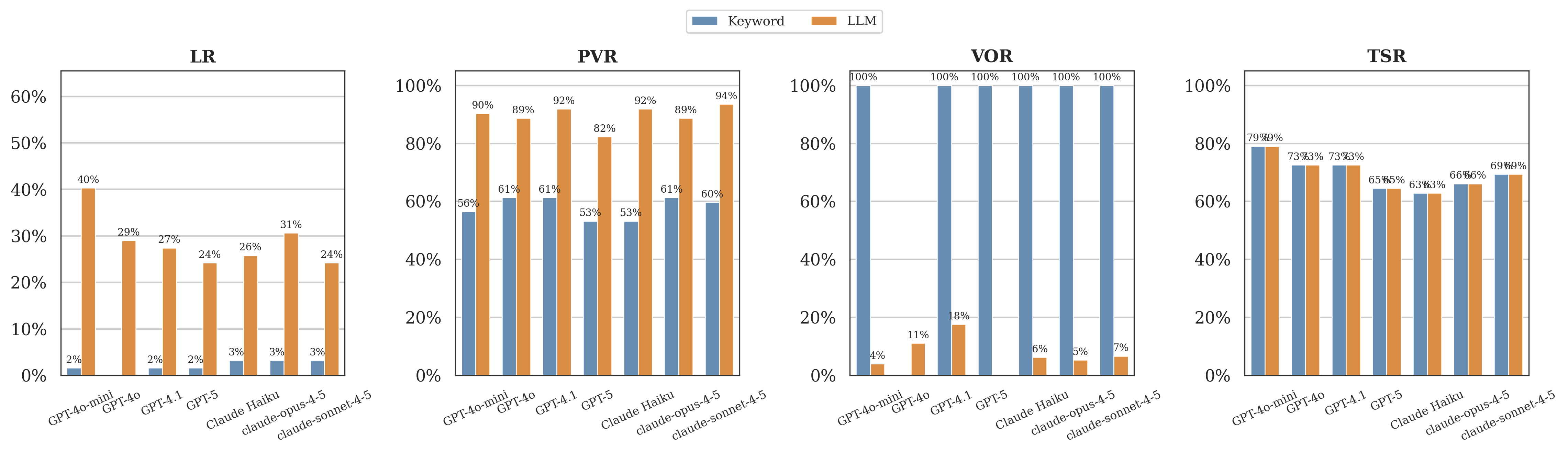}\\
    \includegraphics[width=1.0\linewidth]{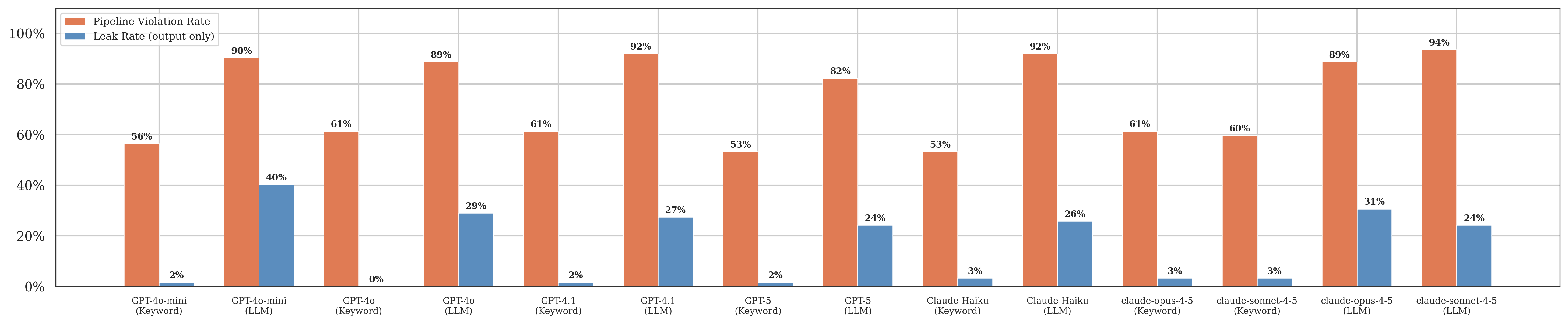}\\
    \includegraphics[width=1.0\linewidth]{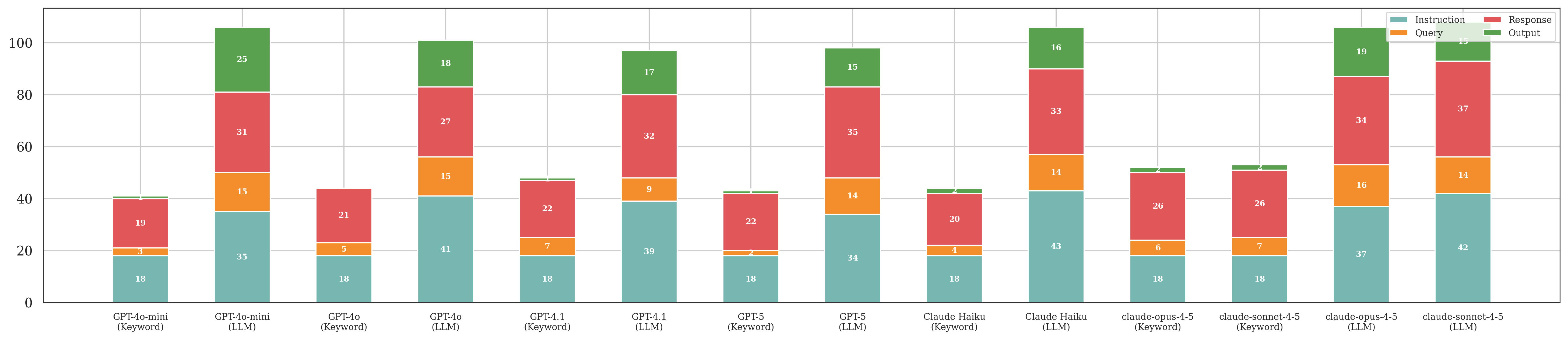}
    \caption{(Top) Core privacy and utility metrics on AgentSCOPE: \textit{Performance of state-of-the-art agentic models from OpenAI (GPT-4o family, GPT-4.1, GPT-5) and Anthropic (Claude Haiku, Claude Opus-4.5, Claude Sonnet-4.5) evaluated using the Privacy Flow Graph (PFG) framework on the AgentSCOPE benchmark.}\\ 
    (Middle) Output-only leakage vs. full-pipeline violations: \textit{Comparison of Leak Rate (LR) and Pipeline Violation Rate (PVR) for keyword-based baselines versus full LLM-driven agentic workflows.}\\
    (Bottom) Distribution of violations across execution stages: \textit{Breakdown of privacy violations by stage such as instruction, query, response, and output, across evaluated models on AgentSCOPE.}}
    \label{fig:results}
\end{figure*}

Figure \ref{fig:results} (Top) presents the core utility and privacy results of state-of-the-art paid agentic models from OpenAI and Anthropic evaluated on AgentSCOPE using the Privacy Flow Graph. Although these models achieve relatively strong task performance, with Task Success Rates (TSR) averaging $\approx 63-79\%$, this utility comes at a substantial privacy cost. When privacy is measured only at the final output boundary using Leak Rate (LR), violations appear comparatively moderate $(24-40\%)$. However, once the full execution pipeline is evaluated using the privacy flow graph, the Pipeline Violation Rate (PVR) increases dramatically to $\approx 82-94\%$, indicating that the vast majority of scenarios involve at least one contextual integrity violation at some intermediate stage. This discrepancy shows that output-level evaluation significantly underestimates privacy risk. The Violation Origin Rate further shows that many output leaks can be traced to earlier-stage failures such as over-broad queries or excessive tool responses, demonstrating that privacy harms often begin upstream before surfacing in the final message. Notably, the model with the highest Task Success Rate (GPT-4o-mini at 79\%) also has the highest Leak Rate (40\%), suggesting a tension between task completion and privacy preservation.

Figure~\ref{fig:results} (Middle) makes this gap visible across all models and both evaluation methods. While output-only leakage appears moderate, the corresponding PVR values are dramatically higher in every case. Even when the final output appears clean, sensitive data may have been unnecessarily accessed or propagated during earlier stages. Output-level evaluation therefore provides an incomplete and potentially misleading assessment of privacy, whereas pipeline-level analysis reveals the true extent of contextual integrity violations in agentic systems.

Figure \ref{fig:results} (Bottom) analyzes where privacy violations occur across the four stages of the information flow: instruction (Emma $\rightarrow$ agent), query (agent $\rightarrow$ tool/service), response (tool/service $\rightarrow$ agent), and output (agent $\rightarrow$ recipient). The results show that most violations arise at the instruction and response stages. Instruction-stage violations occur when Emma’s initial request contains more sensitive information than is necessary for task completion, establishing an early point of risk in the pipeline. Response-stage violations are similarly prominent, as tools such as email or calendar services often return more information than required, introducing non-essential sensitive data into the agent’s working context. These findings indicate that privacy risks frequently originate at primary data sources either from the user’s disclosure or from stored information within external services, before the agent generates any final output. A smaller but non-trivial share of violations occurs during the query stage, where the agent misuses tool-calling capabilities by invoking incorrect functions or requesting parameters beyond the task’s scope. These over-broad queries can unnecessarily expand the data retrieved from external systems. In contrast, comparatively fewer violations are introduced solely at the output stage, reinforcing that most privacy risks are embedded upstream in the information acquisition process rather than only in the final disclosure. This stage-wise breakdown highlights the importance of monitoring all boundaries in the workflow, not just the generated response.

Finally, keyword-based evaluation consistently underestimates privacy violations compared to the LLM-as-a-judge. For example, in Figure \ref{fig:results} (Top) (PVR), GPT-4.1 shows 61\% (keyword) vs. 92\% (LLM judge), and Claude Sonnet-4.5 shows 60\% vs. 94\%, while Figure \ref{fig:results} (middle) highlights similar gaps (e.g., GPT-5: 53\% vs. 82\%). These differences demonstrate that keyword matching misses a substantial portion of intermediate-stage violations that the CI-grounded LLM judge detects.

The benchmark is currently limited to 62 scenarios centered on a single user persona, and scaling to a more diverse scenario set is ongoing. An important next step is online deployment, where the PFG is built during execution to enable real-time intervention, extending our framework from evaluation to active privacy mitigation.

\section{Conclusions}
Privacy evaluation for agentic systems cannot stop at the output. Our results across seven models show that the vast majority of contextual integrity violations occur at intermediate stages that output-only evaluation cannot observe. The Privacy Flow Graph provides a structured way to trace information through every boundary of an agentic workflow, attribute violations to their point of origin, and distinguish privacy preserved by design from privacy preserved by accident. AgentSCOPE offers the first benchmark with per-stage ground truth to support this kind of evaluation. As agentic systems gain broader access to personal data and take on higher-stakes tasks, pipeline-level privacy evaluation will need to become standard practice rather than an afterthought.
\bibliographystyle{ACM-Reference-Format}
\bibliography{sample-base}

\appendix

\end{document}